\pdfoutput=1

\documentclass[useAMS,usenatbib]{mnras}

\usepackage{float}
\usepackage{mathtools}
\usepackage[table]{xcolor}

\usepackage[T1]{fontenc}

\usepackage{graphicx}	
\usepackage{amsmath}	
\usepackage{amssymb}	
\usepackage{multicol}        
\usepackage{bm}		
\usepackage{pdflscape}	

\usepackage[T1]{fontenc}

\title[\textit{The initial mass function of M82 disk SSCs}]{The cluster initial mass function of the M82 disk Super Star Clusters}

\author[Cuevas-Otahola et al.]{
B. Cuevas-Otahola,$^{1,2}$\thanks{E-mail: bolivia@inaoep.mx}
Y. D. Mayya,$^{2}$
J. Arriaga-Hernández,$^{3}$
I. Puerari,$^{2}$ and
\newauthor
G. Bruzual $^{1}$
\\
$^{1}$Instituto de Radioastronom\'ia y Astrof\'isica, Universidad Nacional Aut\'onoma de M\'exico, Morelia, Michoac\'an, 58089, M\'exico\\
$^{2}$Instituto Nacional de Astrof\'isica, \'Optica y Electr\'onica, 72840 Puebla, Mexico\\
$^{3}$Mathematics Department-FCFM, Benemérita Universidad Autónoma de Puebla, Puebla, 72570, Puebla, Mexico\\
}

\date{Accepted 2023 August 26. Received 2023 August 23; in original form 2022 November 11}

\pubyear{2023}

\begin{document}
\label{firstpage}
\pagerange{\pageref{firstpage}--\pageref{lastpage}}
\maketitle

\begin{abstract}
The presence of a population of a large number ($\sim$400) of almost coeval (100--300~Myr) super star clusters (SSCs) in the
disk of M82 offers an opportunity to construct the Cluster Initial Mass Function (CIMF) from the observed present-day Cluster Mass Function (CMF). We carry out the dynamical and
photometric evolution of the CMF assuming the clusters move in circular orbits under the gravitational potential of the host galaxy using the semi-analytical simulation code EMACSS. We explore power-law and log-normal functions for the CIMFs, and populate the clusters in the disk assuming uniform,
power-law, and exponential radial distribution functions. We find that the observed CMF is
best produced by a CIMF that is power-law in form with an index of 1.8, for a power-law radial distribution function. More importantly, we establish that the observed turn-over in the
present-day CMF is the result of
observational incompleteness rather than due to dynamically induced effects, or an intrinsically log-normal CIMF, as
was proposed for the fossil starburst region B of this galaxy. Our simulations naturally reproduce the mass-radius relation observed for a sub-sample of M82 SSCs.
\end{abstract}

\begin{keywords}
galaxies: clusters: general --  globular clusters: general -- catalogues
\end{keywords}




\section{Introduction}
Star clusters are fundamental units of star formation in galaxies \citep{LadaLada2003}. The gravitational potential of the stars and gas 
within the cluster volume and the kinetic energy of stars need to be in Virial equilibrium at all times for the long-term survival of 
star clusters. All clusters may not be bound at birth, while others become unbound due to the expulsion of gas enclosed within the 
cluster volume due to multiple supernova expulsions in the cluster \citep{GoodwinBastian2006}.
\citet{Adamo2017} find that these effects can decrease the number 
density of clusters by a factor larger than 3 for ages between 10 and 100 Myr. The survival of clusters for periods longer than around 
100~Myr depends on the gravitational potential of their host galaxies, as well as their location in the galaxy \citep{Vesperini2001,Fall&Zhang}. The most massive and compact clusters are expected to live for a Hubble time, in which case they could be 
progenitors of present-day Globular Clusters (GCs)
\citep[e.g.][]{Kruijssen2015,Cuevas2020b}. This subset of massive and 
compact clusters are generally referred to as Super Star Clusters (SSCs). For the purpose of this work, we define SSCs as young (age 
$<$1~Gyr) clusters having masses $M>10^4\, M_\odot$  and half-light radii $R_h<$~10 pc.

One of the ways to test the hypothesis that the SSCs are the progenitors of GCs is to study their mass functions, after accounting for the dynamically-induced destruction processes
\citep{Kroupa1995}. Cluster Mass Functions (CMFs) are usually obtained from the luminosity functions, after applying age-dependent mass-to-light ratios using Population Synthesis Models  \citep[e.g.][]{BruzualCharlot2003}. The CMFs followed by star clusters at birth, i.e., the Cluster Initial Mass Function (CIMF) can be obtained from studies of young clusters (age $\sim$~1—10~Myr), which are yet to experience long-term dynamical effects caused by the tidal effects of the parent galaxy. Hence the observed CMF of these young clusters is usually assumed to be identical to the CIMF as the corrections for mass-loss during stellar evolution and early dynamical evolution can be neglected over such a short timescale \citep{Kroupa2002}. Examples of such studies can be found in galaxies such as M51 \citep{Bik2003}, Antennae \citep{ZhangFall1999,Benedict2002}, LMC \citep{deGrijs2008}, the starburst galaxies NGC 3310 and NGC 6745 \citep{deGrijs2003} and in the nuclear region of M82  \citep{McCrady2007,Mayyacat}. The CIMF thus obtained has been found to be well described by a power-law function
$\frac{dN}{dm} \propto m^{-\alpha}$ with index $\alpha=2$ between $10^3\, M_\odot$ and $ 10^6\, M_\odot$ \citep{Krumholz}. 
\citet{Larsen2009} suggested a truncation of the mass function at the high-mass end, with the CMF well described by the \citet{Schechter1976} distribution, used to describe the luminosity function (LF) of galaxies. The break is represented by the truncation mass $M_*$ known as the cut-off mass in the CIMF context. $M_*$ is thought to give insights into the local environment as a function of the galaxy type  \citep{Larsen2009,Portegiesrev}, with starburst galaxies having larger $M_*$ values on average than the normal spirals and irregulars, which could be suggesting a dependence on the environment pressure, with starburst galaxies having higher pressures \citep{Sun2015}. More recently, several works have found a relation between the Star Formation Rate surface density $\Sigma_{\rm SFR}$ and $M_*$, in several galaxies: M83 \citep{Adamo2015}, M31 \citep{Johnson2017}, M51 \citep{Messa2018}, and M33 \citep{Wainer2022}. In particular, \citet{Johnson2017} 
reported a relation 
$M_* \propto ⟨\Sigma_{\rm SFR}⟩^{\sim 1.1}$, which they suggest also applies to Globular Clusters, giving insights into the current formation models.

If SSCs are progenitors of GCs, both should have similar mass functions at birth. However, the luminosity function of GCs, which is 
considered as a proxy for mass function, is found to be log-normal \citep{Jordan2007,Lomeli2022}. 
These differences are thought to be due to the nature of tidal forces that over long period of time selectively destroy low mass 
clusters \citep{Vesperini2001,Fall&Zhang}. The change from power-law to log-normal form happens gradually as the clusters orbit under the gravitational potential of their parent galaxy. In this context, study of clusters that have ages comparable to their orbital periods (around 100~Myr or slightly older; intermediate-age clusters, henceforth) provide us an opportunity to catch them when this change is happening.

The disk of the nearby starburst galaxy M82 contains a rich population of intermediate-age SSCs, first reported by \citet{deGrijs2001} using the HST/WFPC2 camera images of the "fossil starburst" region B \citep{Oconnel1978}. \citet{Mayya2006} carried out a detailed photometric and chemical population synthesis model to reproduce the observed
spectroscopic, photometric and dynamical properties of the disk of M82, and proposed a disk-wide star-formation event around 500~Myr
ago with a duration of around 300~Myr, with no star-formation in the disk in the last 100~Myr. The fly-by encounter of M82 with the 
members of the M81 group around a gigayear ago \citep{Yun1994} is the most likely trigger for the disk-wide star formation. 
The presence of a rich population of luminous AGB stars throughout the disk also favours an intense post-interaction star formation 
event \citep{Davidge2008}. These latter studies also noted the absence of Red Supergiants in the disk, which supports the absence of star
formation over the last 10 to 50~Myr.
Intense events of star-formation are accompanied by the formation of SSCs, and hence the whole disk, not just the region B, is expected 
to contain a rich population of SSCs. 
The wider field of view of the HST/ACS camera mosaic images allowed \citet{Mayyacat}
\citep[see also][]{Lim2013} to establish
the presence of SSCs in the entire disk of M82. The ages for the disk population reported in follow-up spectroscopic and photometric studies range between 100~Myr to around 1~Gyr as detailed below. \citet{deGrijs2001} analysed the optical WFPC2 BVI and NICMOS JH photometric data of SSCs in region B and estimated an age of 
around 600~Myr for these clusters. \citet{deGrijs2003a} reported a slightly older age of ~1~Gyr from a reanalysis of the same dataset. \citet{Smith2007} carried out the HST/STIS optical spectroscopy of some of these clusters in region B, obtaining slightly younger ages of 350~Myr. \citet{Konst2009} carried out multi-object spectroscopy of 49 SSCs, obtaining a mean spectroscopic age of $\sim$~150 Myr for the disk SSCs. 
\citet{RodriguezMerino2011} analysed multi-band Spectral Energy Distributions (SEDs) over spatial scales of 180~pc, 
obtaining an age range of 100--450~Myr for the disk populations. 
The ages obtained in this latter study for the SSCs in region B agree well with that reported by \citet{Smith2007}. It should be noted that the ages obtained using photometric colours covering only optical and infrared wavelength range are affected 
by age-reddening degeneracy, which unfortunately introduces larger uncertainties in M82 as compared to the normal disk galaxies given 
its large inclination angle and dusty morphology. 
On the other hand, spectroscopic ages are not affected by age-reddening degeneracy, and hence are more reliable.
The relatively small spread in the spectroscopically derived ages of 100--350~Myr makes the M82 disk cluster sample
an ideal sample to study to understand the evolution of the CMF from the power-law for young SSCs to log-normal for the old GCs. Analysing the SSCs in region B, \citet{deGrijs2003a} obtained a log-normal CMF with a turn-over mass $2\times 10^5\, M_\odot$, that was well above the 50\% completeness limit. They suggested that, if clusters form following an initial power-law mass function, its transformation to a log-normal shape could happen in $\sim$1 Gyr due to the calculated
 30~Myr disruption timescale in M82 disk.
In a follow-up study, \citet{deGrijs2005} used analytical prescriptions based on previous N-body results by \citet{BaumgardtMakino2003} to conclude that the observed log-normal CMF was not the result of transformation, instead clusters in the M82-B region were formed with a log-normal CIMF. The high-density environments in which these clusters were formed are thought to be the reason for the log-normal CIMF. On the other hand, the reported CMFs for young clusters in intense starburst galaxies such as Antennae have a power-law form \citep{Whitmore1999,Benedict2002}. This brings to an interesting question as to why the CIMF in M82-B region is different from that in other starburst regions?

In this work, we analyze the CIMF of the cluster sample for the disk of M82, with a special interest to test whether the M82 disk SSCs obey a log-normal CIMF as found by \citet{deGrijs2005}, or instead a power-law CIMF like in other starburst systems. The analysis is motivated by the availability of a fast numerical tool, called Evolve Me a Cluster of StarS \citep[EMACSS,][]{AlexanderGieles2014}, which allows the dynamical evolution of star clusters under the influence of the gravitation potential of a 
parent galaxy, assuming the clusters have circular orbits around the center of the galaxy. 
We used this tool to construct the CIMF that is consistent with the present-day CMF, assuming a uniform age of 100~Myr for all SSCs.
The chosen initial set of cluster parameters cover a wide range of initial masses
and radii, placed at different galactocentric distances that simulate the presently observed radial distribution of SSCs, and
obeying power-law and log-normal CIMFs. 
In \S\ref{Sec:Models}, we describe the initial conditions of the cluster sample, as well as their corresponding mass and half-light radius evolution. We include a prescription for the observational biases in order to compare the observed and simulated clusters properly. We summarize our conclusions in \S\ref{Sec:conclu}

\section{Observed Mass function for M82 disk SSCs}\label{Sec:Obs_data}

We use the mass function obtained by 
\citet{Mayyacat} as the current-day CMF for M82 disk SSCs. This function is based on a sample of 393 SSCs in the disk of this galaxy, detected in the Hubble Space Telescope (HST) Advanced Camera for Surveys (ACS) archive images from the Hubble Heritage Team \citep{Mutchler2007}. This dataset contains information in the F435W, F555W and F814W bands, with a spatial sampling of 0.05 arcsec pixel$^{-1}$, corresponding to 0.88 pc pixel$^{-1}$ at the distance of M82 \citep[3.63 Mpc,][]{Freedman}. For the whole sample, photometric masses were derived from simple stellar population models, assuming that the sample of disk SSCs is coeval.
This assumption of coevality of the disk population is supported by the modeling of the disk properties by 
\citet{Mayya2006} and
the subsequent observational confirmation of a relatively small spread of the spectroscopically-determined ages of disk SSCs \citep{Smith2007,Konst2009}. \citet{Mayyacat} obtained the photometric masses and reddening using a uniform age of 100 Myr and 
\citet{Cardelli1989} extinction 
curve with Rv=3.1, with the reddening values determined as the average of $F435W-F555W$ and $F555W-F814W$  
excesses over the SSP colours for the assumed age. Use of an older age would increase the mass of a cluster of given magnitude due to the 
increase of mass-to-light ratio with age, but would decrease the mass due to a lower inferred color excess, with the net effect of
increase of the mass by less than a factor of two if all clusters are as old as 1~Gyr, instead of the assumed 100~Myr.
\citet{Mayyacat} found that the resulting CMF follows a power-law with a slope $\alpha$=1.5 between $10^4-10^6\, M_\odot$. 
The CMF shows a turnover at $\sim$10$^4\, M_\odot$, which they associated to the incompleteness in the detection of lower mass clusters.

\section{Cluster simulations and dynamical models}{\label{Sec:Models}}

We use Monte Carlo simulations to populate the disk of M82 with cluster populations obeying five different CIMFs. Star clusters are distributed at different galactocentric distances for three assumed functional forms. For a cluster
of given mass its half-mass radius is chosen so that the total population of clusters follows previously-defined functional forms of their mean density distributions. Each cluster is then evolved under the gravitational potential of a typical late-type disk galaxy for 100~Myr. The details of each of these functions are explained below.

The 100 Myr for the cluster population fixed in our simulations is based on the 
small spread in observed ages around this value for M82 disk clusters as 
discussed in detail in the introduction.  At $t=100$~Myr, stellar evolution and early dynamical processes like two-body relaxation and disk shocks \citep{Fall&Zhang} are already at play.
We use EMACSS to simulate how these effects affect the evolution of the clusters. EMACSS treats dynamical evolution in terms of the relaxation time and the flow of energy normalised to the initial cluster energy. Tidal fields are included assuming an isothermal halo profile. 
In the following subsections we describe the initial conditions used in our simulations.

\begin{figure*}
\begin{center}
\includegraphics[width=\textwidth]{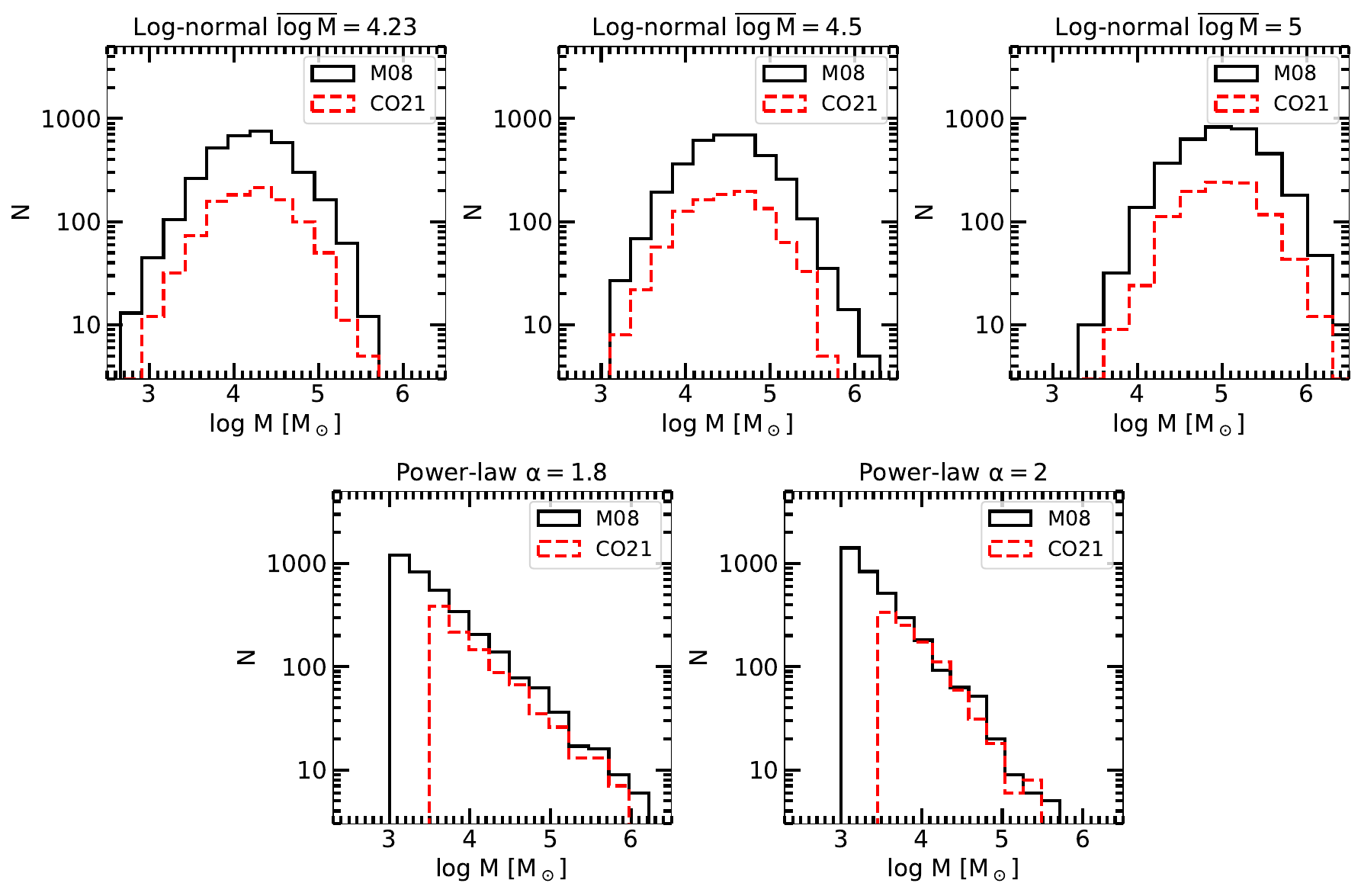}
\caption{Simulated CIMF to reproduce the observed samples in \citet{Cuevas2020b} (CO21) (red dashed line) and \citep{Mayyacat} (M08) (black solid line)
following log-normal functions (upper panels), with $\overline{\log(\rm M)}$
equal to 4.23, 4.5, 5 $\rm M_\odot$ from left to right, and power-law CIMF (bottom panels) with $\alpha=1.8$ (left) and $\alpha=2$ (right). The simulated CIMF have 1000 and 3500 points in order to reproduce the observed CMF in CO21 and M08, respectively.}
\label{Fig:ini_mass}
\end{center}
\end{figure*}

\begin{figure*}
\begin{center}
\includegraphics[width=0.8\textwidth]{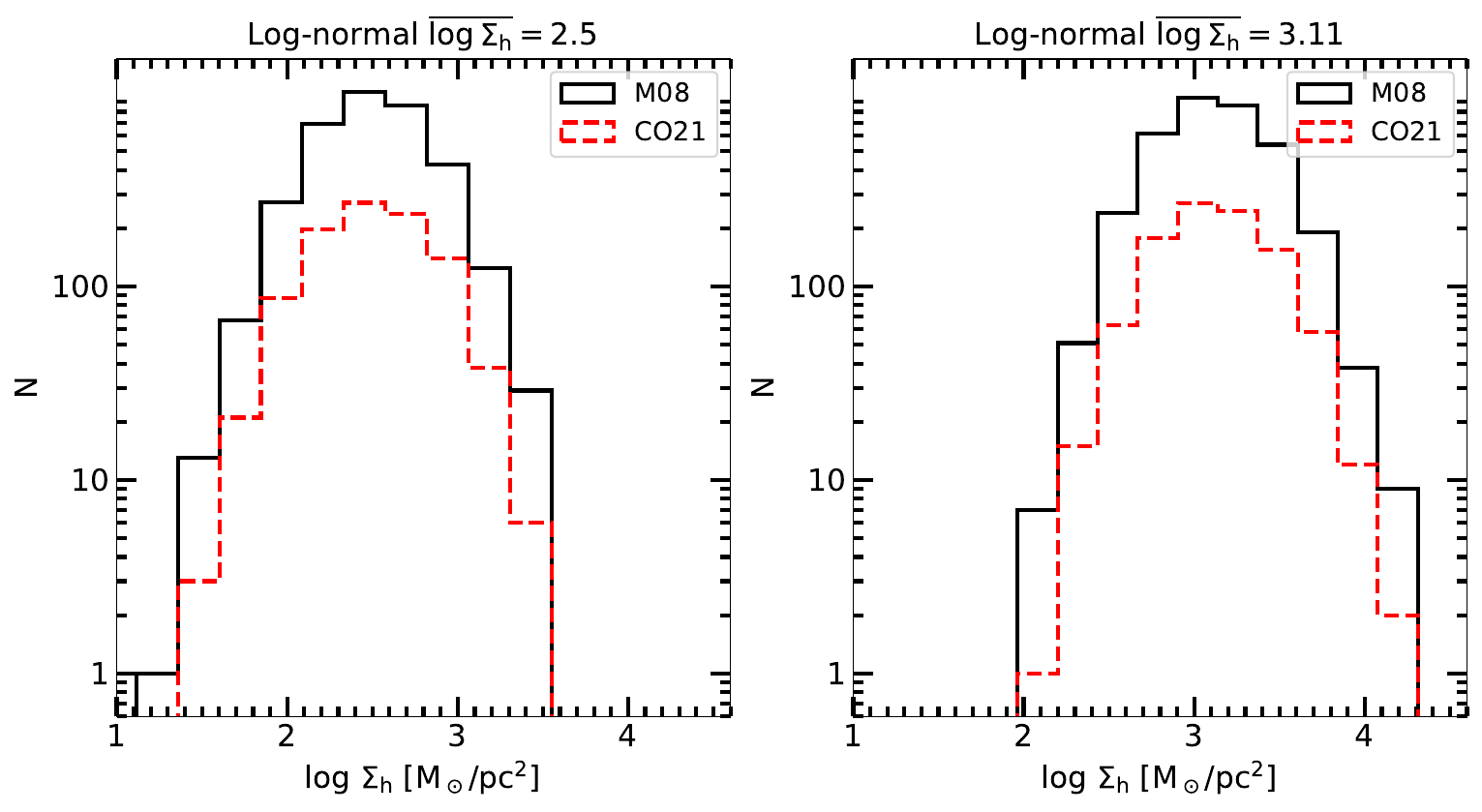}
\caption{Simulated cluster initial $\rm \Sigma_h$ distribution to reproduce the observed samples in \citet{Cuevas2020b} (CO21) (red dashed line) and \citet{Mayyacat} (M08) (black solid line)
following a power-law form with $\overline{\log{\rm \Sigma_h}}$
equal to 2.5 $\rm M_\odot/pc^2$ in the left panel and 3.11 $\rm M_\odot/pc^2$ in the right panel. The simulated distributions have 1000 and 3500 points, in order to reproduce the observed CMF in CO21 and M08, respectively.}
\label{Fig:ini_Ih}
\end{center}
\end{figure*}

\begin{figure*}
\begin{center}
\includegraphics[width=\textwidth]{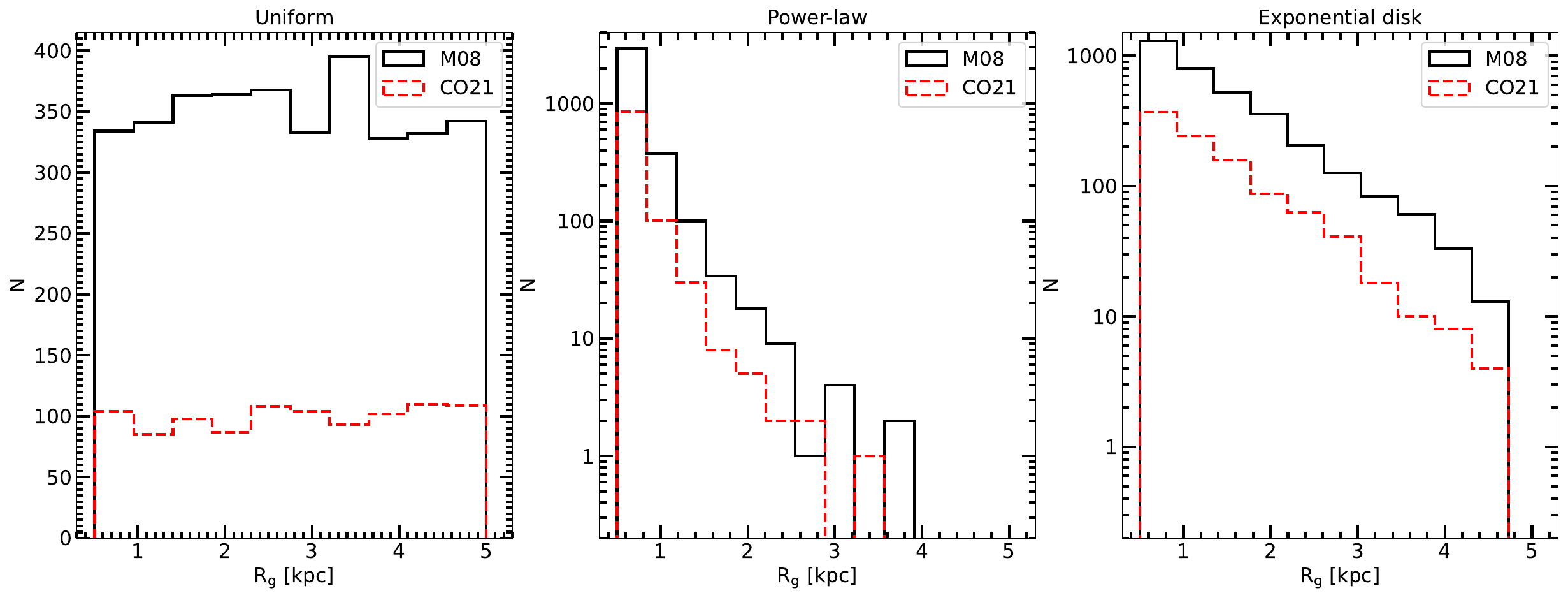}
\caption{Simulated cluster initial $\rm R_g$ distributions to reproduce the observed samples in \citet{Cuevas2020b} (CO21) (red dashed line) and \citet{Mayyacat} (M08) (black solid line), drawn from uniform (left), power-law (center), and exponential disk (right) functions.  The simulated distributions have 1000 and 3500 points in order to reproduce the observed CMF in CO21 and M08, respectively.}
\label{Fig:rg_ini_Ih}
\end{center}
\end{figure*}

\subsection{Model CIMFs}

Following the work by \citet{deGrijs2005}, we explore three sets of log-normal \citep{Vesperini1998,Vesperini2000,Vesperini2001} and two sets of power-law \citep{Fall&Zhang} CIMFs, in order to reproduce the observed CMF in  \citet{Mayyacat} (hereinafter M08). We draw the log-normal CIMFs from log-normal distributions centred at $10^{4.23}$, $10^{4.5}$, and $10^{5}$ $\rm M_\odot$ with $\rm \sigma \log (M/M_\odot)$=0.33 using Monte Carlo techniques.  
The power-law CIMFs are drawn from the functions $\rm (100\, M/10^{4.5}\, M_\odot)^{-1.8}$ and $\rm (100\, M/10^{4.5}\, M_\odot)^{-2}$. We set the power-law low-mass limit to $log(M/M_\odot)>3$. It may be noted that clusters with masses lower than this limit would be
below the detection limits. Moreover, these relatively low-mass clusters 
become unbound much before the current age of 100 Myr for the population. 
The resulting CIMFs are constituted by n=3500 objects each and are shown in Fig.~\ref{Fig:ini_mass} in black solid lines. 

Most M82 clusters are in the process of expansion, with a few having expanded up to more than 20\% of the Jacobi radius (see Fig. 9 in CO21). At the present rate of expansion, very few clusters would survive for more than 500~Myr.  For this reason, we discard the possibility of ages as old as 1 Gyr for the M82 
cluster population as was reported in some studies \citep{deGrijs2001}, and end our
simulations at 100~Myr.

\subsection{Model initial radius distributions}

\citet{Cuevas2020,Cuevas2020b}
carried out a structural analysis of a subset of 99 SSCs from the 
\citet{Mayyacat}
M82 disk SSC sample. This subset represents the bright (massive) end of the original sample, restricted mainly by the relatively high surface 
brightness of the disk and the amount of crowding of SSCs in M82. The sample however is representative of the original sample for clusters massive than $3\times 10^4 M_\odot$
The analysis was carried out using the code \textsc{nProFit} \citep{CuevasnProfit}, developed to exploit the
capability of the HST in the characterization of core and halos of the SSCs in nearby galaxies. 
\textsc{nProFit} extracts the surface brightness profiles of a given sample of star clusters, fits empirical and dynamical models (either Moffat-EFF, King, or Wilson) to the extracted profiles, and returns dynamically relevant parameters such as the half-light 
radius ($\rm R_h$), the total mass ($M$) and the velocity dispersion of stars $\sigma$, the latter two parameters calculated using the mass-to-light 
ratio that corresponds to the assumed age. We used the mass-to-light ratio value of 0.11346 in the F555W band, which corresponds to the value for 100 Myr, with the Kroupa IMF \citep{KroupaIMF} in the BC03 models \citep{BruzualCharlot2003}.

The distribution of stellar half-mass  surface densities ($\Sigma_h$) for a subsample of 99 clusters was presented in  \citet{Cuevas2020b} (hereinafter CO21). The authors used EMACSS code to derive the initial distribution of surface densities, finding that clusters  are in a state of expansion if $\log(I_h)<4 \frac{L_\odot}{pc^2}$ or equivalently, $\Sigma_h < 1135 \frac{M_\odot}{pc^2}$ for the adopted mass-to-light ratio. The distribution of the initial mean surface density of these expanding clusters followed a log-normal form (see Figure 7 in CO21). Hence, as a starting point, we assume the mean surface density distribution obtained for the 99 SSCs in CO21 holds for the entire sample of disk SSCs and use this log-normal form for the distribution of the initial mean surface density. We assume two log-normal  $\rm \Sigma_h$ distributions centred at $10^{2.5}$ and $10^{3.11}$ $\rm M_\odot /pc^2$ and $\rm \sigma \log (\Sigma_h/M_\odot pc^{-2})$=0.33, shown in Fig.~\ref{Fig:ini_Ih}  in red dashed lines, from which we draw the initial $\rm R_h$ distributions.

\subsection{Distribution of galactocentric distances of simulated clusters}
In order to represent the effect of tidal forces on the cluster evolution,  in the simulations we place each SSC at a galactocentric radius ($\rm R_g$) drawn from three sets of initial $\rm R_g$ distributions:  uniform, power-law, and exponential disk functions.  In CO21 we found that 9\% of SSCs are likely to become Globular Clusters (GCs). Considering this, we have chosen the index of our power-law $\rm R_g$ distribution to be -4.5, which reproduces the Milky Way GC $\rm R_g$ distribution \citep{Baumgardt1998}. 
We also consider an exponential disk function following \citet{Mayya2009}, who report such a function with a scale-length of 1 kpc in the V-band. 
The initial $\rm R_g$ distributions for CO21 and M08 are shown in Fig.~\ref{Fig:rg_ini_Ih} in red dashed and black solid lines, respectively. 
The lower and upper $\rm R_g$ limits of these distributions are 0.5 and 5 kpc, which correspond to the present-day $\rm R_g$ values for the observed sample of SSCs (M08).
We summarized the initial conditions to reproduce the CIMF in M08 and its subsample CO21 in Tab. \ref{tab:tabla1}.

\begin{table*}
\begin{center}
\caption{Initial conditions (to reproduce CMF in Paper I)}
\label{tab:tabla1}
\begin{tabular}{ccllllll}
\hline
Run & CIMF & $f(R_g)$ & $\overline{\log M}$ & $\overline{\log \Sigma_h}$ & $\overline{\log M}$ & $\overline{\log \Sigma_h}$ \\ 
 & & &  $\rm M_\odot$ & $\rm M_\odot/pc^2$ & $\rm M_\odot$ & $\rm M_\odot/pc^2$  \\
 (1) & (2) & (3) & (4) & (5) & (6) & (7)\\
\hline
1  &  log-normal  &  uniform  &  4.23  &  3.11  &  4.23  &  3.11  \\
2  &  log-normal  &  power-law  &  4.23  &  3.11  &  4.23  &  3.11  \\
3  &  log-normal  &  exponential  &  4.23  &  3.11  &  4.23  &  3.11  \\
4  &  log-normal  &  uniform  &  4.23  &  2.5  &  4.23  &  2.5  \\
5  &  log-normal  &  power-law  &  4.23  &  2.5  &  4.23  &  2.5  \\
6  &  log-normal  &  exponential  &  4.23  &  2.5  &  4.23  &  2.5  \\
7  &  power-law $\alpha=$2  &  uniform  &  3.82  &  3.11  &  3.43  &  3.11  \\
8  &  power-law $\alpha=$2  &  power-law  &  3.82  &  3.11  &  3.43  &  3.11  \\
9  &  power-law $\alpha=$2  &  exponential  &  3.82  &  3.11  &  3.43  &  3.11  \\
10  &  power-law $\alpha=$2  &  uniform  &  3.82  &  2.5  &  3.43  &  2.5  \\
11  &  power-law $\alpha=$2  &  power-law  &  3.82  &  2.5  &  3.43  &  2.5  \\
12  &  power-law $\alpha=$2  &  exponential  &  3.82  &  2.5  &  3.43  &  2.5  \\
13  &  log-normal  &  uniform  &  4.5  &  3.11  &  4.5  &  3.11  \\
14  &  log-normal  &  power-law  &  4.5  &  3.11  &  4.5  &  3.11  \\
15  &  log-normal  &  exponential  &  4.5  &  3.11  &  4.5  &  3.11  \\
16  &  log-normal  &  uniform  &  4.5  &  2.5  &  4.5  &  2.5  \\
17  &  log-normal  &  power-law  &  4.5  &  2.5  &  4.5  &  2.5  \\
18  &  log-normal  &  exponential  &  4.5  &  2.5  &  4.5  &  2.5  \\
19  &  power-law $\alpha=$1.8  &  uniform  &  3.86  &  3.11  &  3.54  &  3.11  \\
20  &  power-law $\alpha=$1.8  &  power-law  &  3.86  &  3.11  &  3.54  &  3.11  \\
21  &  power-law $\alpha=$1.8  &  exponential  &  3.86  &  3.11  &  3.54  &  3.11  \\
22  &  power-law $\alpha=$1.8  &  uniform  &  3.86  &  2.5  &  3.54  &  2.5  \\
23  &  power-law $\alpha=$1.8  &  power-law  &  3.86  &  2.5  &  3.54  &  2.5  \\
24  &  power-law $\alpha=$1.8  &  exponential  &  3.86  &  2.5  &  3.54  &  2.5  \\
25  &  log-normal  &  uniform  &  5  &  3.11  &  5  &  3.11  \\
26  &  log-normal  &  power-law  &  5  &  3.11  &  5  &  3.11  \\
27  &  log-normal  &  exponential  &  5  &  3.11  &  5  &  3.11  \\
28  &  log-normal  &  uniform  &  5  &  2.5  &  5  &  2.5  \\
29  &  log-normal  &  power-law  &  5  &  2.5  &  5  &  2.5  \\
30  &  log-normal  &  exponential  &  5  &  2.5  &  5  &  2.5  \\
\hline
\end{tabular}
\hfill\parbox[t]{\textwidth}{Description of the columns: (1) Run label, (2) Cluster initial mass function for the simulations, (3) Function used to generate the $\rm R_g$ initial distributions. (4) Mean mass of the initial clusters mass distributions for CO21. (5) Mean of the initial surface brightness distributions for CO21.(6) Mean mass of the initial clusters mass distributions for M08. (7) Mean of the initial surface brightness distributions for M08.}
\end{center}
\end{table*}

\subsection{Dynamical evolution of the simulated clusters in the gravitational potential of M82}

Several studies of the evolution of the CIMF are based on the photometric evolution of the clusters \citep[e.g.][]{Larsen2002,Bastian2008,Sun2015} using synthetic models of simple stellar populations \citep{BruzualCharlot2003} and
neglecting disruptive dynamical effects that are proven to drive mass loss \citep{Bastian2008}. 

We evolved the simulated clusters of initial mass $M$, and half-mass radius $\rm R_h$, located at the galactocentric distance $\rm R_g$ for a duration of 100~Myr using EMACSS code. The gravitational potential of M82 is described
by a flat rotation curve of 100 km/s velocity 
\citep{Konst2009,Greco2012}. At t = 100 Myr, stellar evolution and early dynamical processes like two-body relaxation and disk shocks 
\citep{Fall&Zhang} are already at play. EMACSS treats dynamical evolution in terms of the relaxation time and the flow of energy normalised to the initial cluster energy. Tidal fields are included assuming an isothermal halo profile, to represent the potential of the host galaxy. 

EMACCS follows stellar evolution (through the evolution of the mean mass of the stellar mass function) and includes a simplified prescription of early dynamical processes that depend on the cluster $\rm R_h$ and considers escapers  \citep{Lamers2010,Gieles2010}. 
A simplified prescription of mass segregation and core collapse is also included in EMACSS.  
The influence of very massive objects, such as stellar black holes and neutron stars, which avoid cluster collapse and lead to a later core expansion \citep{Mackey2008} is considered in EMACSS using a rough approximation,
based on the fact that the evolution of $\rm R_h$, unlike the core radius, does not depend on the retention of black holes \citep{Lutzgendorf2013,BreenHeggie2013}.  The approximation performed by EMACSS is reliable for times up to twice the core collapse time, with the size of clusters beyond that timescale being even smaller \citep{AlexanderGieles2014}.
At the selected age ($t=100$~Myr), the approximation is reliable. To prove this statement, we have selected the initial conditions of the densest simulated cluster, surviving for 100 Myr, and we recall the expressions for the relaxation time $t_{\rm rh}=N/8 \ln N t_{\rm cr}$ where the crossing time $t_{\rm cr}=1/  \sqrt{G \rho}$ \citep{SpitzerHart1971}.  Such a cluster has an initial density of 10$^{4.7}$~$M_\odot / pc^3$, resulting in $t_{\rm rh}=318$ Myr.  The time of core collapse $t_{\rm cc}$ is given in terms of $t_{\rm rh}$, which in the case of clusters constituted by equal mass stars is of the order of 15--20 $t_{\rm rh}$ \citep{Fujii2014}, whereas for clusters with mass distributed following a mass function is 0.2  ~$t_{\rm rh}$ \citep{PortegiesMcMillan2002} which yields $t_{\rm cc}$=64 Myr. Hence, EMACSS would give unreliable results for such a dense cluster for $t>128$ Myr, which is above the time we are evolving the clusters.  Regarding external perturbations, EMACSS treatment of tidal fields is based on circular orbits, assuming a constant circular velocity, and an isothermal halo potential. We assume a flat rotation curve with a velocity of 100 km s$^{-1}$, following the values reported for M82 \citep{Konst2009,Greco2012}. 
The tidal effects in EMACSS are set by defining the galaxy velocity and the galactocentric radius.   
Along with the tidal effects, EMACSS allows to include an experimental prescription for dynamical friction (the sinking of clusters towards the galactic center). However, such an effect can be neglected at intermediate ages. For example, the most massive simulated cluster surviving up to 100 Myr has a mass of 10$^{6.16}$~$M_\odot$.  From Eq.~(7.26) in \citet{Galdynbook}, for the latter cluster we have a time of dynamical friction $t_{\rm df}$ of 460~Myr, which is almost 5 times larger than the evolution time analysed. Additionally to the previously mentioned effects, interaction with giant molecular clouds (GMCs) and spiral arms are disruptive effects with considerable effects on clusters with masses below 10$^4$~$M_\odot$ in the solar neighbourhood \citep{LamersGieles}.
We bear in mind that EMACSS does not have  a prescription for GMC interactions, hence, such effect is not accounted in the simulations. In order to understand the effects of GMC interactions, in CO21, we re-wrote Eq. (4)  in \citet{LamersGieles} for cluster disruption due to GMCs. Both effects, dynamical friction and interaction with GMCs have a dependence on the cluster mass and $\rm R_h$, as well as on environmental properties such as the cloud densities.  Assuming Milky Way-like cloud properties, we have a disruption time due to interaction with GMCs  for the least massive cluster in CO21 surviving for 100 Myr (M=288~$M_\odot$ and $\rm R_h=0.38$~pc) of 71 Gyr, and for the spiral arms interaction a disruption time of 574 Gyr, following Eq.~(5) of \citet{LamersGieles}.

The starting number of clusters is determined a-posteriori so that $\sim$400 clusters (i.e. close to the observed number of SSCs) survive after evolution 
for 100~Myr. We found this initial number to be N=3500. The exact number defines the normalization factor only and not the shape of the CMF. It may be recalled that the $\rm R_h$ for a cluster of mass $M$ is drawn to satisfy the $\Sigma_h$ distribution obtained for the subset of 99 SSCs,
which is slightly (by around 20\%) biased towards massive SSCs). There is no reason to believe that the $\Sigma_h$ distribution for the entire sample would be different from that for the subset of 99 SSCs. 
In order to guard against any bias introduced due to this assumption, we also ran the simulations to produce the CMF for the subset of 99 SSCs. The initial distributions for this subset are shown in Figures 1 to 3 by red lines.

The sixty sets of initial conditions described above  produce clusters with very diverse dynamical configurations.
The Jacobi radius ($\rm R_j$) takes into account the galactic properties, and the $\rm R_h / \rm R_j$ ratio provides information on the Roche volume filled by a cluster. We recall that $\rm R_j$ is a proxy of the tidal field strength and is given as a function of the cluster angular frequency $\Omega$, and the cluster mass as $R_j^3\propto \frac{M}{\Omega^2}$ \citep{King_dyn}.
To understand the initial dynamical configuration of the simulated clusters, we compute the initial $\rm R_h /\rm R_j$ and find that SSCs simulated using power-law CIMFs are, on average, more embedded within their $\rm R_j$, whereas those drawn from log-normal CIMFs show a larger fraction of  tidally-limited clusters.  
For the log-normal mass distributions, $\sim$65\% of the clusters are initially tidally-limited, whereas for the power-law initial mass distributions the corresponding fraction is  $\sim$53\%.

\subsection{Dynamical evolution of clusters using EMACSS}

\begin{figure*}
\begin{center}
\includegraphics[width=0.8\textwidth]{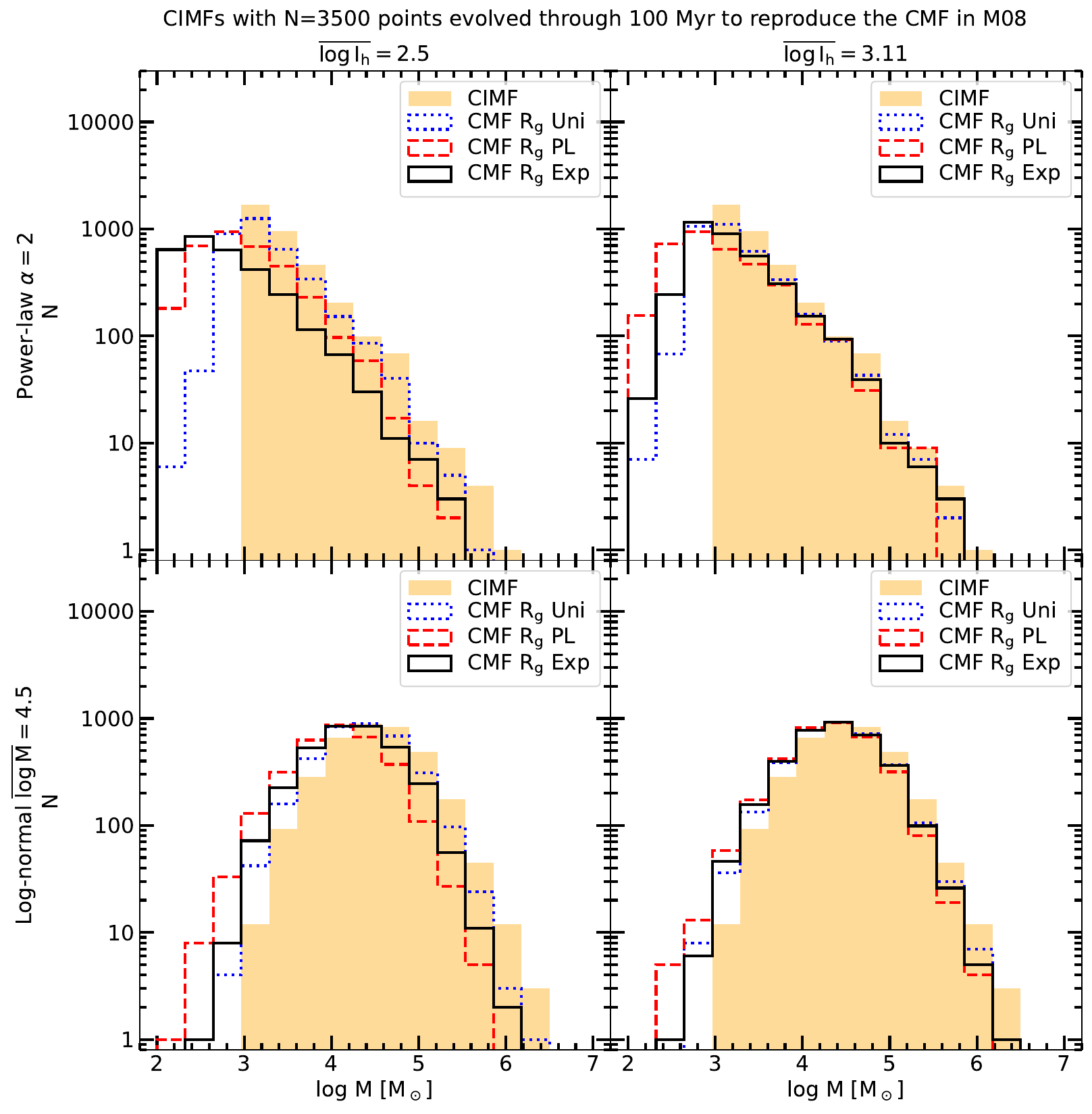}
\caption{CIMFs evolved through 100 Myr using the semi-analytical evolution code EMACSS. In the left panels, we show CIMFs evolving assuming lower surface density distributions, whereas in the right panel, the denser ones. In the top panels we show the evolved CIMFs drawn from  power-law distributions with $\alpha=2$ (runs 7 to 12), whereas in the  bottom panels we show the evolved CIMFs drawn from log-normal distributions with $\overline{\log M=4.5}$ (Runs 13 to 18) under uniform tidal fields (uniform $\rm R_g$ distributions) and exponential and power-law $\rm R_g$.  The evolved CMFs are compared with the CIMFs in Fig. \ref{Fig:ini_mass} (yellow histogram).}
\label{Fig:evol_mass_mayya}
\end{center}
\end{figure*}

\begin{figure}
\begin{center}
\includegraphics[width=\columnwidth]{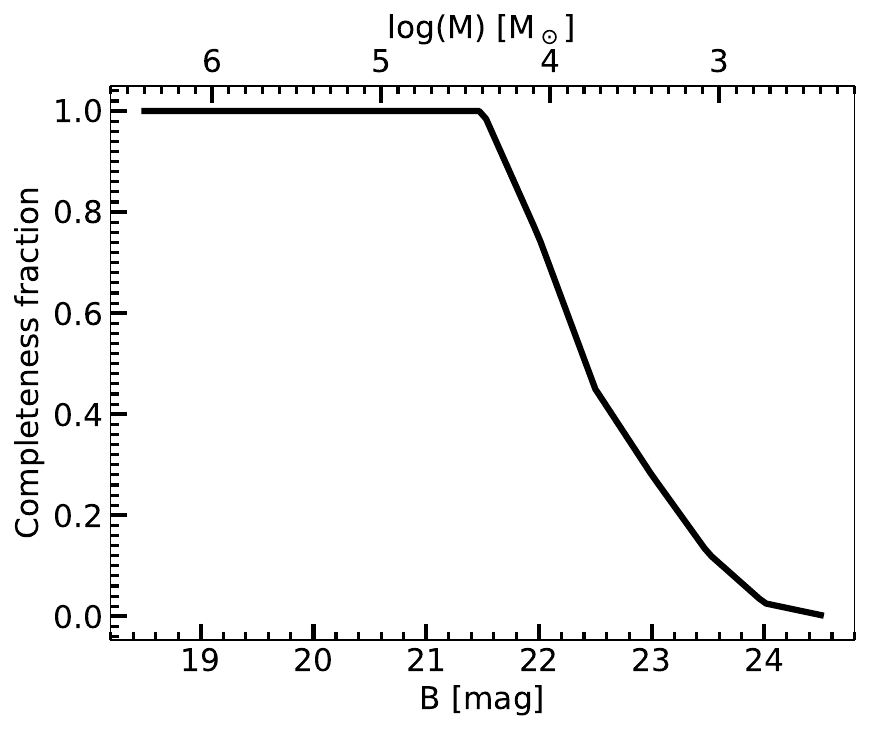}
\caption{Completeness function used in this work, derived from  \citet{Mayyacat} (M08)}
\label{Fig:comp_func}
\end{center}
\end{figure}

In CO21, we have shown that under the galactic field of M82, massive clusters at 100 Myr evolve resembling the evolution of isolated clusters. Less dense clusters lying within 2 kpc in the disk, suffer from larger mass-loss than isolated clusters. Clusters experience an initial expansion phase, which is stalled once the cluster reaches its tidal limit. Such a limit is given by the $\frac{R_h}{Rj}$ ratio \citep{AlexanderGieles2014}, reaching values larger than 0.1 for tidally limited clusters.  The early gas expulsion has a critical role in the evolution of clusters, which is crucial in explaining the formation scenario  of clusters out of virial equilibrium during the early phases of evolution. Such non--virial clusters could have been born that way or such configurations could arise from the violent gas expulsion \citep{Larsen2009}. Since EMACSS does not include a prescription for gas expulsion, we consider the former scenario.

Throughout this paper we assume that stars form following the  \citet{KroupaIMF} initial mass function (IMF), with lower, upper, and mean masses of 0.1, 100, and 0.6337  $M_\odot$, respectively.

\subsubsection{CIMF evolution}\label{Sec:mass_ev}

In Fig. \ref{Fig:evol_mass_mayya} we show the evolved mass distributions at 100 Myr for two CIMFs to illustrate the mass evolution in terms of the three galactocentric distributions (power-law, exponential disk, and uniform) and two surface density distributions (low and high density). We show for illustration purposes only power-law with $\alpha=2.0)$ and log-normal CIMF with $\overline {\rm logM}=4.5$, considering that the overall behavior is  similar for all power-law as well as for every log-normal CIMFs. We show in the upper and bottom panels the evolved power-law ($\alpha=2.0$) and log-normal ($\overline {\rm logM}=4.5$) distributions, respectively. In the left panels, we show the less dense whereas in the right panels the denser evolved distributions. Each panel shows the distributions evolving under three different galactocentric initial functions. In the figure, we qualitatively compare the evolved CIMFs with their corresponding CIMF, intended to reproduce the CMF in M08.  

The following two main conclusions can be drawn from the simulations: 1. The evolution does not change the form of the CIMF, i.e. both the power-law and log-normal CIMFs retain their initial forms for all assumed $\rm R_g$ distributions for cluster masses above $10^3 M_\odot$, the minimum mass in simulations.
2. The present-day CMFs have a low-mass tail as compared to the CIMFs. In fact, there is a systematic shift towards lower masses by around 0.2--0.7 dex over the entire mass range, with larger shifts for lower mass clusters. These effects are more pronounced for the lower-density clusters (left panels). The shift for the standard exponential distribution of $\rm R_g$ (solid black lines) is intermediate between uniform (dotted blue) and power-law (dotted red) $\rm R_g$ distributions. This is expected as the power-law form overpopulates clusters in the inner regions where the tidal effects are maximum, as compared to the exponential form. On the other hand, the uniform distribution populates most clusters in the external parts where tidal effects are minimum. For comparison, the mean $\rm R_g$ for the simulated
samples after 100~Myr of evolution are 0.6, 1.1, and 2.7~kpc, respectively for the power-law, exponential, and uniform distribution of $\rm R_g$ values.

It may be recalled that the initial $\rm R_h$ for each cluster is chosen in such a way that the whole population satisfies the initial $\Sigma_h$ distribution, where $\Sigma_h = M_{cl}/\pi*R_h^2/2$. The $\Sigma_h$ distribution is based on the subsample of 99 SSCs studied in CO21. In Appendix sections \ref{Sec:ApMass} and \ref{Sec:ApRh}, we demonstrate that the above conclusions are valid for the mass function for the subsample of 99 clusters also.

\begin{figure*}
\begin{center}
\includegraphics[width=0.87\textwidth]{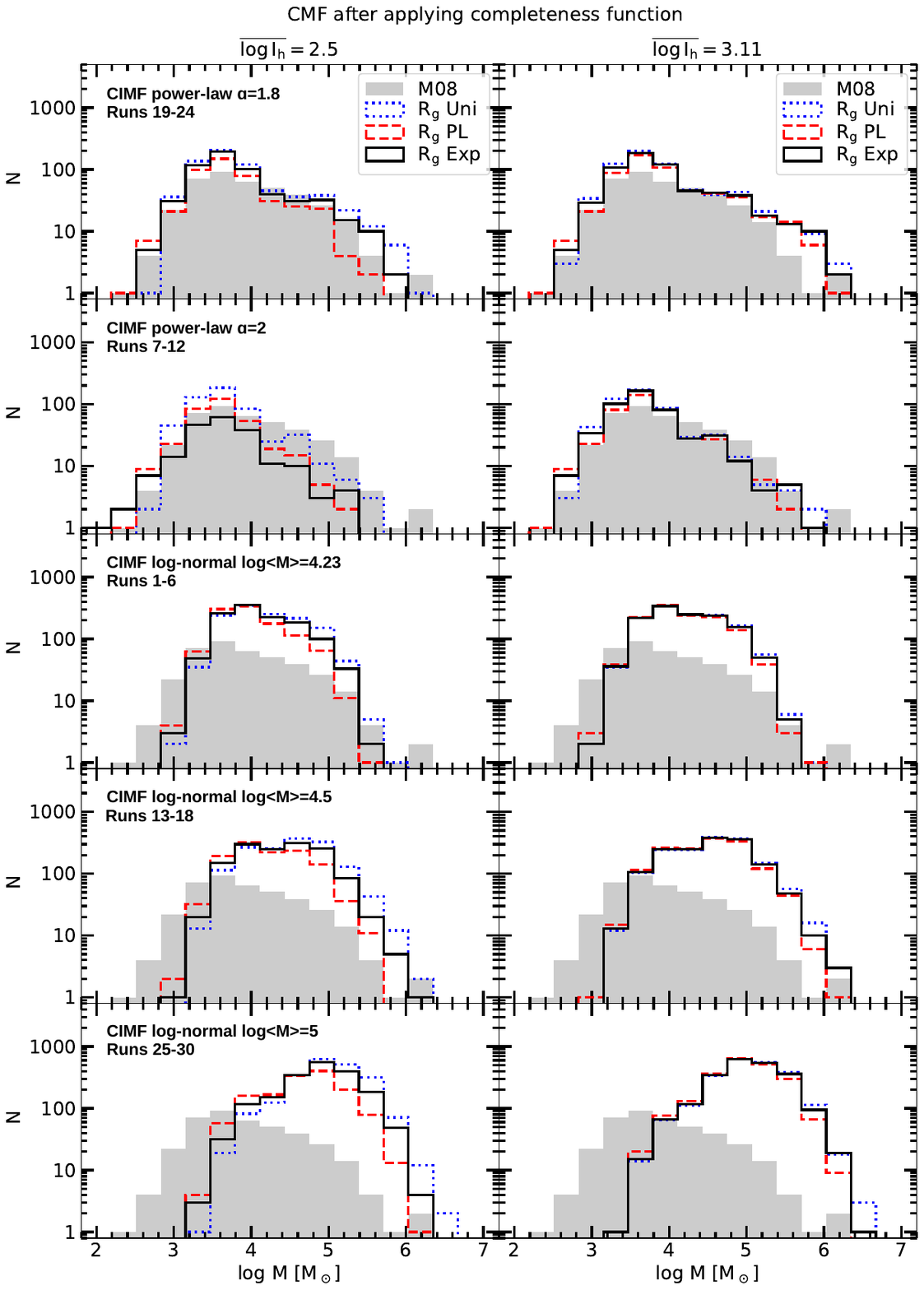}
\caption{CIMFs evolved through 100 Myr using the semi-analytical evolution code EMACSS, and applying the completeness function in Fig. \ref{Fig:comp_func}. In the left panels, we show CIMFs evolving assuming lower surface density distributions, whereas in the right panels, the denser ones. The first
and second panels show evolved CIMFs drawn from power-law distributions (Runs 19 to 24 and 7 to 12), and from the middle to the bottom panels evolved CIMFs drawn from log-normal distributions (Runs 1 to 6, 13 to 19, and 25 to 30), under uniform tidal fields (uniform $\rm R_g$ distributions) and exponential and power-law $\rm R_g$.  The evolved CIMFs are compared with the observed CMF in  M08 (gray histogram).}
\label{Fig:evolve_bias_mayya}
\end{center}
\end{figure*}

\begin{figure*}
\begin{center}
\includegraphics[width=0.87\textwidth]{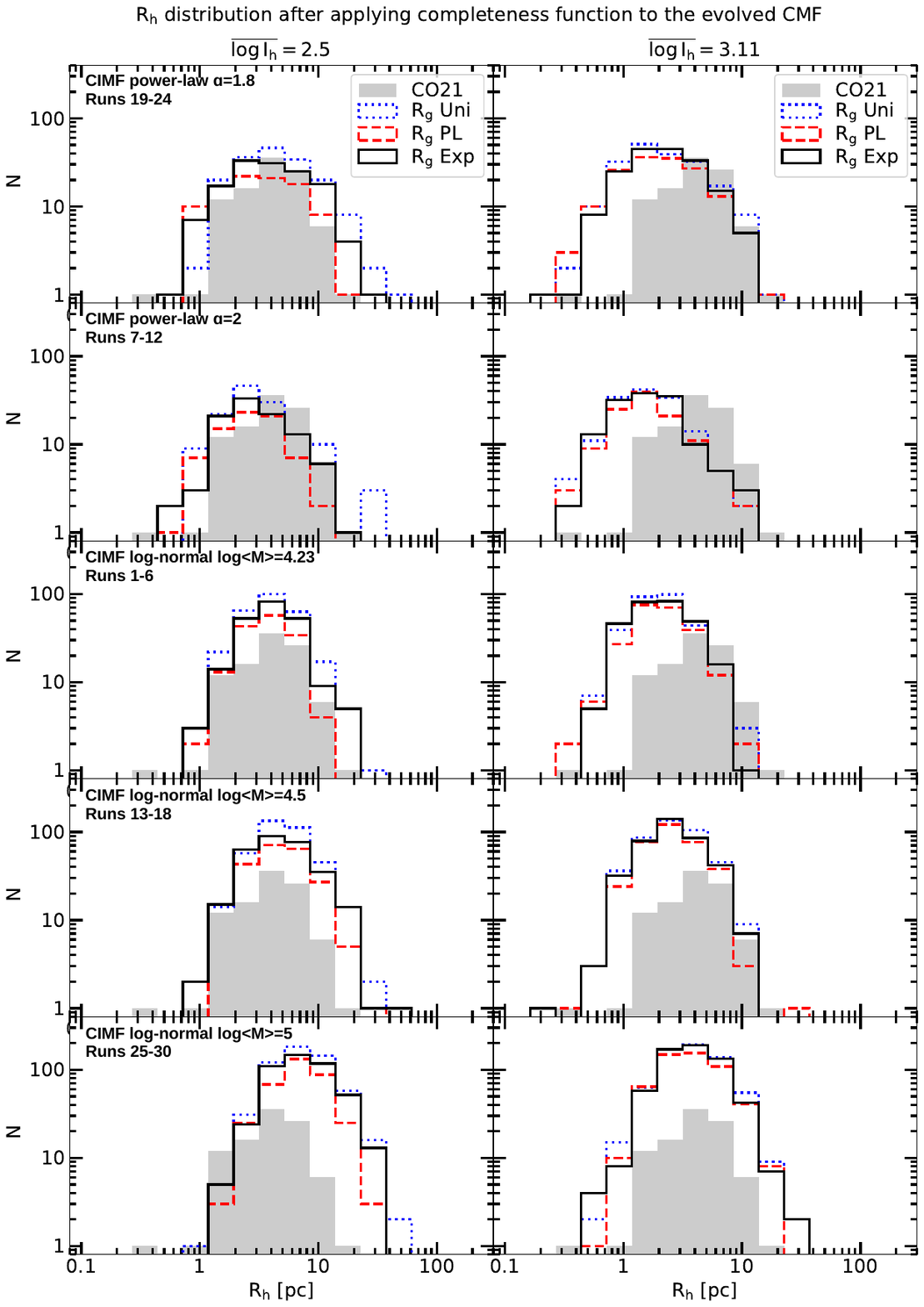}
\caption{$\rm R_h$ distribution evolved through 100 Myr using the semi-analytical evolution code EMACSS, and applying the completeness function in Fig. \ref{Fig:comp_func}. In the left panels, we show the $\rm R_h$ distributions evolving assuming lower surface density distributions, whereas in the right panels, the denser ones. The first
and second panels show the $\rm R_h$ distributions evolved, with masses following CIMFs drawn from power-law distributions (Runs 19 to 24, and 7 to 12), and from the middle to the bottom panels evolved CIMFs drawn from log-normal distributions (Runs 1 to 6, 13 to 19, and 25 to 30), under uniform tidal fields (uniform $\rm R_g$ distributions) and exponential and power-law $\rm R_g$.  The evolved $\rm R_h$ distributions are compared with the observed ones in  CO21 (gray histogram).}
\label{Fig:evol_rad_bias}
\end{center}
\end{figure*}

\subsubsection{Correction of simulated CIMF for observational incompleteness}{\label{Sec:obs_eff}}

\begin{table}
\begin{center}
\caption{Mean masses and half-light radii of the evolved simulated distributions, and their p-values obtained from the Kolmogorov-Smirnov test (K-S).}
\label{tab:tabla2}
\begin{tabular}{lllllll}
\hline
Run & $\overline{\log \rm M}$ & p-M & $\overline{\log \rm M}$ & p-M  & $\overline{\rm R_h}$ & p-R \\
 & $\rm M_\odot$ & & $\rm M_\odot$ & & pc & \\
 (1) & (2) & (3) & (4) & (5) & (6) & (7)\\
\hline
1 & 4.23 & 1.33e-15 & 4.43 & 2.38e-08 & 2.02 & 2.89e-16 \\
2 & 4.19 & 0.00e+00 & 4.40 & 1.92e-07 & 1.96 & 1.65e-16 \\
3 & 4.22 & 2.14e-21 & 4.44 & 4.36e-08 & 2.05 & 4.34e-16 \\
4 & 4.20 & 3.44e-15 & 4.41 & 5.61e-08 & 3.99 & 0.82 \\
5 & 4.01 & 7.14e-09 & 4.28 & 1.20e-03 & 3.41 & 0.03 \\
6 & 4.13 & 1.61e-14 & 4.37 & 3.61e-06 & 3.95 & 0.66 \\
7 & 3.70 & 2.82e-10 & 4.17 & 0.11 & 1.58 & 3.22e-21 \\
8 & 3.73 & 6.31e-07 & 4.19 & 0.18 & 1.53 & 6.18e-16 \\
9 & 3.72 & 3.09e-08 & 4.19 & 0.07 & 1.52 & 5.84e-20 \\
10 & 3.67 & 1.00e-10 & 4.11 & 0.09 & 2.98 & 1.28e-03 \\
11 & 3.63 & 1.03e-11 & 4.02 & 0.15 & 2.69 & 1.26e-04 \\
12 & 3.64 & 3.13e-07 & 4.08 & 0.11 & 2.76 & 1.58e-04 \\
13 & 4.55 & 1.10e-57 & 4.66 & 1.27e-13 & 2.67 & 9.56e-09 \\
14 & 4.49 & 3.10e-51 & 4.62 & 2.86e-12 & 2.48 & 4.87e-11 \\
15 & 4.52 & 7.77e-16 & 4.65 & 5.81e-14 & 2.63 & 1.67e-09 \\
16 & 4.50 & 1.07e-52 & 4.62 & 6.99e-12 & 4.88 & 0.02 \\
17 & 4.22 & 2.22e-15 & 4.42 & 3.26e-08 & 4.69 & 0.09 \\
18 & 4.39 & 4.44e-15 & 4.52 & 2.15e-10 & 4.75 & 0.12 \\
19 & 3.90 & 1.42e-02 & 4.44 & 1.67e-03 & 2.06 & 1.40e-12 \\
20 & 3.92 & 6.51e-02 & 4.46 & 2.01e-03 & 1.86 & 6.17e-13 \\
21 & 3.91 & 3.36e-02 & 4.50 & 4.06e-04 & 2.02 & 1.85e-11 \\
22 & 3.86 & 8.40e-04 & 4.43 & 3.81e-03 & 4.63 & 0.09 \\
23 & 3.74 & 0.128 & 4.15 & 0.58 & 3.69 & 0.17 \\
24 & 3.81 & 0.103 & 4.30 & 0.03 & 4.07 & 0.33 \\
25 & 5.04 & 1.48e-158 & 5.01 & 1.58e-32 & 3.89 & 0.58 \\
26 & 4.97 & 8.88e-16 & 4.95 & 4.90e-30 & 3.72 & 0.20 \\
27 & 5.02 & 1.78e-154 & 4.99 & 1.72e-32 & 3.74 & 0.22 \\
28 & 4.98 & 1.67e-15 & 4.95 & 1.18e-30 & 7.50 & 7.55e-14 \\
29 & 4.68 & 1.93e-76 & 4.72 & 9.06e-17 & 6.73 & 2.82e-11 \\
30 & 4.87 & 1.89e-15 & 4.86 & 2.42e-24 & 7.37 & 1.23e-12 \\
\hline
\end{tabular}
\hfill\parbox[t]{\columnwidth}{Description of the columns: (1) Run label, (2)  Mean mass of the evolved clusters CIMF intended to reproduce the observed CMF in M08, (3) Parameter p of the Kolmogorov-Smirnoff test, obtained when comparing the observed CMF in M08 and the simulated CIMFs, (4) Mean mass of the evolved CIMFs intended to reproduce the observed CMF in CO21,  (5) Parameter p of the Kolmogorov-Smirnoff test, obtained when comparing the observed CMF in CO21 and the simulated CIMFs, (6)  Mean half-light radius ($\rm R_h$) of the evolved clusters $\rm R_h$ distributions intended to reproduce the observed distribution in CO21, (7) Parameter p of the Kolmogorov-Smirnoff test, obtained when comparing the observed $\rm R_h$ distributions and simulated $\rm R_h$ distributions.} 
\end{center}
\end{table}

Before comparing the results of the simulated clusters to the observed CMF, the bias 
caused due to incompleteness of the observed sample has to be incorporated to the 
simulated CMFs. M08 have presented in their Figure~5 the completeness correction curve as a function of the observed magnitude for the sample of SSCs from which
the CMF we analyse here was obtained. For the sake of clarity, we show the completenes function used in this section, in Figure \ref{Fig:comp_func}, as a function of the B-magnitude from which the masses in M08 are computed, and for illustration purposes we also show the completeness as a function of the clusters' masses. We used the mass-to-light ratio for 100~Myr to transform their completeness
function as a function simulated mass. 

In Figure~\ref{Fig:evolve_bias_mayya}, we compare the observed CMF from M08 with the CMF for our simulated clusters for all the runs in Table \ref{tab:tabla1} after applying the effects due to incompleteness of the observed CMF. The observed CMF from M08 is shown in grayscale, whereas the three $\rm R_g$ distributions used in this work are shown by lines of different types. The models for lower density clusters are shown on the left plots, whereas the higher density models are shown to the right.  It can be inferred by a simple look at the plots that the power-law models are better match to the observed distribution
of masses. On the other hand, none of the three log-normal distributions correctly produce the observed range of masses. The log-normal models that reproduce the correct number of clusters at the two extreme ends produces several orders of
magnitude excess number of clusters of intermediate masses as compared to the observed numbers.

In order to quantitatively compare the observed distributions with the simulated ones, we carry out the statistical non-parametric Kolmogorov-Smirnov test (K-S) \citep{KStest} to determine whether the simulated and observed distributions are comparable. In Tab. \ref{tab:tabla2}, we show the mean mass values for M08 (column 2) and CO21 (column 4), along with the $\rm R_h$ mean values for CO21 (column 6) for all runs, labelled in column 1, along with the   p-parameters of the K-S test (columns 3, 5 and 7), which allow us to reject or not the null hypothesis that two compared distributions are drawn from the same parent distribution.  If p is below 0.01 (1\%), we can reject the null hypothesis, hence we conclude that the compared distributions are different. We observe that all runs performed assuming log-normal CIMFs (runs 1-6, 13-18, 25-30) have p-M values, below 0.01, which supports the arguments laid out before, ruling out initial log-normal distributions to reproduce the observed mass distributions of the M82 disk SSCs sample.

On the other hand, we notice that power-law CIMFs (runs 7-12, 19-24) provide a more accurate representation of the observed samples, most specifically those evolving from less dense distributions (runs 10-12, 22-24) (left second and fourth panels). In particular, the distribution evolving from the relatively shallower CIMF ($\alpha=$1.8) (runs 19-24), display larger p-values than those for the canonical one ($\alpha=$2-case) (runs 7-12). The best-match case is run 23, which corresponds to $\alpha=1.8$ with a power-law $\rm R_g$ distribution. In general, power-law $\rm R_g$ distributions (runs 8,11,20 and 23) provide a marginally better representation of the observed CMF as compared to an exponential $\rm R_g$ distributions runs (9,12,21 and 24). On the other hand, none of the models involving uniform radial distribution represent the observed CMF. Thus in conclusion, we establish that the clusters in M82 disk formed in the same way as in other star-forming galaxies with a power-law, rather than the log-normal form. The simulations favour a slightly shallower power-law index of $\alpha=1.8$, 
as compared to the canonical value of $\alpha=2$.

\begin{figure*}
\begin{center}
\includegraphics[width=0.8\textwidth]{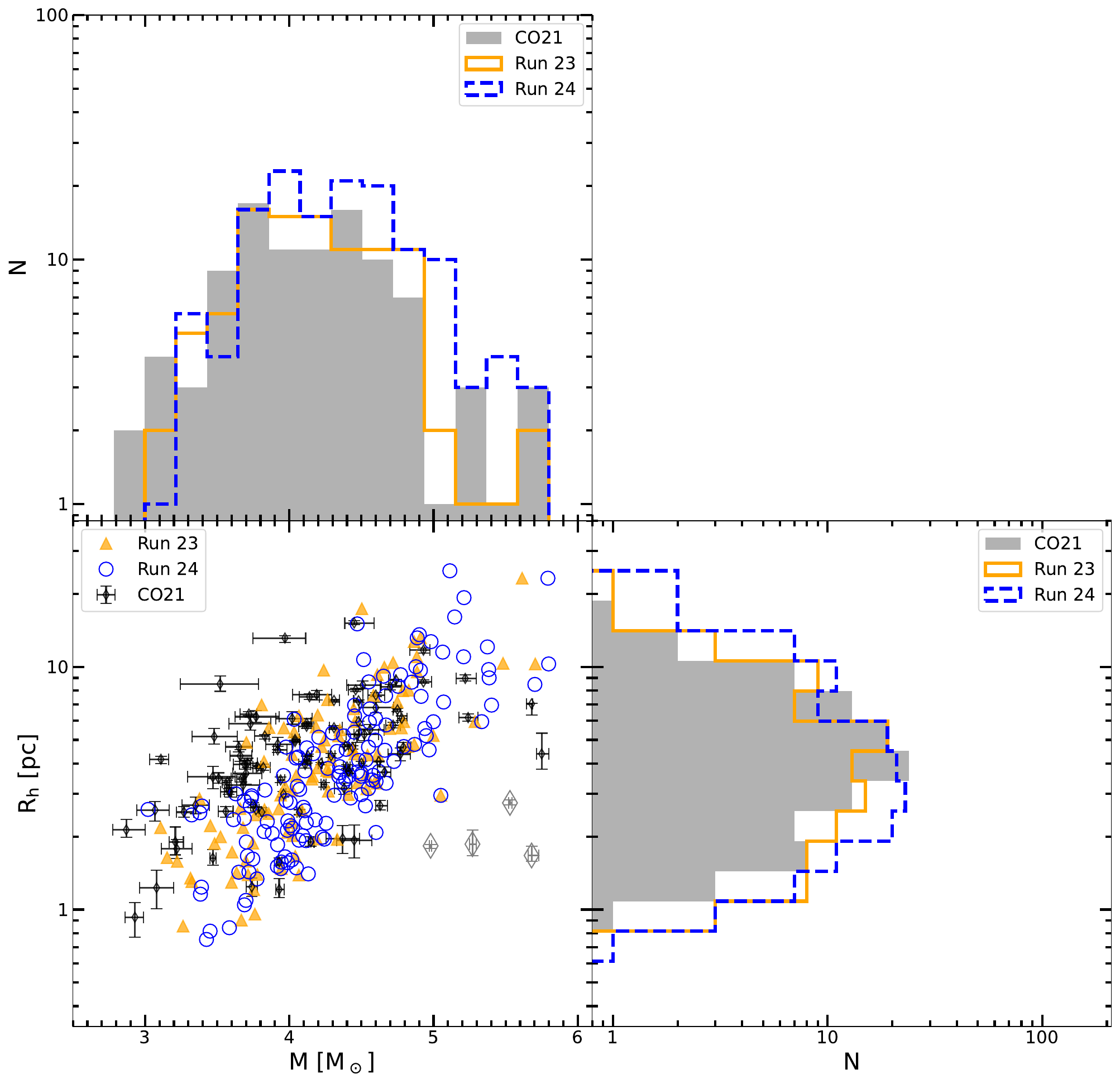}
\caption{(Upper panel) Mass distributions of the evolved simulated clusters in the best runs (23 and 24), evolved through 100 Myr. (Lower left panel) Mass-radius diagram of the best runs, compared with the observed mass-radius relation of M82 disk SSCs, with a small group of compact massive groups in larger and lighter symbols. (Lower right panel) Half-light radius ($\rm R_h$) distributions of the evolved simulated clusters in the best runs.}
\label{Fig:evol_mass_rad}
\end{center}
\end{figure*}

\subsubsection{Comparison of $\rm R_h$ distribution}

In order to be able to study the evolution of the CMF constrained to the mass-radius space, we need to study the effects of the observational biases considered into the M82 disk $\rm R_h$ distribution.  We have studied such effects on the M08 CMF in previous sections. However, as we have described in the introduction, $\rm R_h$ values could be obtained only for a sub-sample of 99 reasonably isolated SSCs (CO21). Nevertheless, we have shown that CO21 is a representative subsample of M08. Hence, we correct the simulated $\rm R_h$ distribution for the observed incompleteness already presented in Fig. \ref{Fig:comp_func} for the total M08 sample. In Fig. \ref{Fig:evol_rad_bias}, we compare the observed distribution of $\rm R_h$ for the sub-sample of 99 SSCs with that for the simulations. We find that as in the mass case, the less dense initial conditions along with power-law distributions (runs 10-12, 22-24) provide distributions comparable to the observed one. However, we draw attention to some p-R values corresponding to log-normal distributions, large enough to accept the null hypothesis, such as that for runs 4 and 6.  Such values indicate that those distributions are drawn from the same distribution, which can be seen in Fig. \ref{Fig:evol_rad_bias}. However, despite the very accurate representation, these runs do not have a good representation for the observed mass distributions, as seen in their very low p-M values. Hence, runs 23 and 24, are the best representations of the observed sample, having p-M and p-R values to conclude they are drawn from the same parent distributions.

\subsubsection{Evolved mass-radius relation}

We have successfully applied observational biases to the CMFs in M08 and its sub-sample in CO21, via the completeness function in Fig. \ref{Fig:comp_func}, derived from Fig. 5 in M08. Hence, we now proceed to analyze the evolution of the mass-radius relation and compare it with the observed trend, reported in CO21. We recall once again that the $\rm R_h$ values are available only for the sub-sample studied by CO21, and hence our comparison of mass-radius relation is restricted 
to CO21 sample only.

In Fig. \ref{Fig:evol_mass_rad}, we compare the results from runs 23 and 24 with the observed mass and $\rm R_h$ distributions of the M82 disk Super Star Clusters (SSCs) in CO21.  Both runs provide good representations of the observed mass-radius relation. However, run 23 (CIMF with index $\alpha=1.8$, power-law $\rm R_g$ distribution and low surface surface density $\overline{\Sigma_h }=2.5$), is more accurate in the less-massive and larger-radii region of the mass-radius diagram. Our simulated MF mean value is 10$^{4.15}\, \rm M_\odot$ which is 0.03 dex above the value in CO21 (10$^{4.12}\, \rm M_\odot$).  We find larger differences in the $\rm R_h$ distribution, with the simulated mean $\rm R_h$ equal to 3.69 pc, whereas the reported one is 4.26 pc. Hence, a shallow CIMF, with low surface density and distributed following an initial power-law $\rm R_g$ distribution as the one in \citet{Baumgardt1998}, reproduces the observed mass-radius relation in the M82 disk SSCs.

Finally, we compare our simulated mass and $\rm R_h$ distributions for the assumed initial conditions in Run 23, with those reported for the observed sample of M82 disk SSCs and find good agreement. There is a small group of four clusters, defined as massive-compact SSCs in CO21, that are not reproduced entirely by our  simulations. It was demonstrated in that work, that this group of clusters corresponds to dense initial conditions, significantly more compact and massive than the initial conditions required to reproduce the rest of clusters in the M82 disk sample. In CO21, these dense models were labeled as M1, M2 and M3.  The group of clusters that obey these dense models are shown by diamond symbols in Fig. \ref{Fig:evol_mass_rad} to distinguish them from the rest of the sample. In  this work, we have only considered density conditions that represent the majority of the cluster sample, corresponding to models M4 to M9 in CO21. We notice in the figure a small group constituted by the less massive and large clusters (below $10^4\, M_\odot$ and above 3 pc), which seem to be slightly  moved from the simulation points. Such a difference could be explained due to the completeness correction used, which is higher at lower masses, introducing larger uncertainties in the low-mass region of the mass distribution, most specifically in the case of low-surface half-mass densities.

\section{Conclusions}{\label{Sec:conclu}}

In this work, we characterized the functional form of the Cluster Initial Mass Function (CIMF) that is consistent with the present-day
(100~Myr) Cluster Mass Function (CMF) of the complete sample of Super Star Clusters in the disk of the late-type galaxy M82. We evolved a population of simulated clusters rotating in circular orbits under the
gravitational potential of M82 using the Evolve Me a Cluster of StarS (EMACSS Alexander et al. 2014) semi-analytical code for a duration of 100~Myr, which is the typical age of the SSCs in the disk of M82. The cluster population was defined by log-normal and power-law forms for the CIMF, and are distributed in the disk of M82 using uniform,
exponential and power-law radial density functions. The initial radius of the clusters were chosen so as to satisfy log-normal distributions of their mean density centered around a low and a high value $(\log \Sigma_h =$ 2.5 and 3.11 $M_\odot/pc^2$, respectively). The simulated CMFs were subjected to the observational incompleteness function and compared with the
observed CMF for the complete sample of SSCs in the disk of M82. Log-normal CIMFs, in general, poorly represent the observed data, with the best fits corresponding to power-law CIMF with an index of $\alpha$=1.8, and $\log \Sigma_h =$ 2.5 $M_\odot/pc^2$. Exponential and Power-law radial distribution functions both represent well the dataset, with the latter distribution giving marginally better fits. The distribution of present-day half-light radius $(R_h)$ for the simulated clusters matches well the observed $\rm R_h$ distribution for a sub-sample of 99 SSCs for which we have measured $\rm R_h$. The latter sub-sample of clusters also follows the observed mass-radius relation. We conclude from the simulations carried out in this work that the clusters in M82 were formed with a power-law CIMF, similar to other starburst systems where such measurements are available. Our simulations demonstrate a turnover of the power-law CMF after 100~Myr of evolution. However, unlike the turnover seen in the Globular Cluster Luminosity Functions, the turnover in our simulations occurs at masses around an order of magnitude lower than the observational limits, for its detection in the M82 cluster sample, which means that the observed turnover is a consequence of completeness instead of an intrinsic phenomenon.

\section*{Acknowledgments}
\addcontentsline{toc}{section}{Acknowledgements}

We thank an anonymous referee whose comments on an earlier version have helped significantly to improve this manuscript.
BCO thanks the DGAPA UNAM Postdoctoral Fellowship Program for the support that enabled her to carry out the work presented here. We also thank CONACyT for the research grants CB-A1-S-25070 (YDM), and CB-2014-240426 (IP).  

\section*{Data Availability}

The data underlying this article are available in the article and in its online supplementary material.

\bsp	
\label{lastpage}
\bibliographystyle{mnras}
\bibliography{bibliografia}
\appendix

\section{Mass and radius evolution}\label{Sec:ApMass}

\begin{figure*}
\begin{center}
\includegraphics[width=0.8\textwidth]{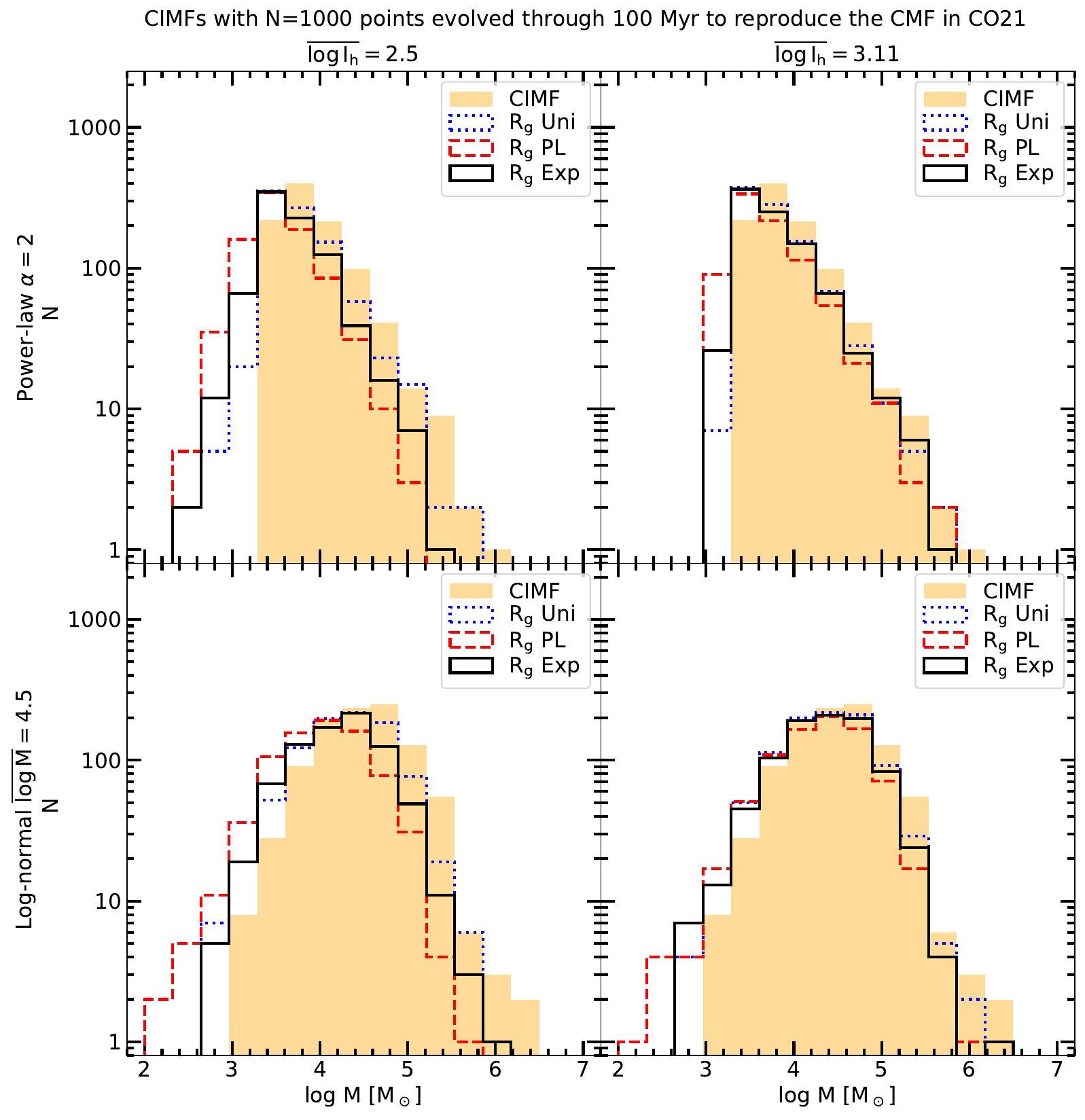}
\caption{CIMFs evolved through 100 Myr using the semi-analytical evolution code EMACSS. In the left panels, we show CIMFs evolving assuming lower surface density distributions, whereas in the right panel, the denser ones. In the top panels we show the evolved CIMFs drawn from  power-law distributions with $\alpha=2$ (runs 7 to 12), whereas in the  bottom panels we show the evolved CIMFs drawn from log-normal distributions with $\overline{\log M=4.5}$ (Runs 13 to 18) under uniform tidal fields (uniform $\rm R_g$ distributions) and exponential and power-law $\rm R_g$.  The evolved CMFs are compared with the CIMFs in Fig. \ref{Fig:ini_mass} (yellow histogram) intended to reproduce the observed CMF in \citet{Cuevas2020b} (CO21)}
\label{Fig:evol_mass}
\end{center}
\end{figure*}

In this section we present an analysis analogous to that in Sec. \ref{Sec:mass_ev}, in this case, for the sub-sample of M08, reported in CO21. In Fig. \ref{Fig:evol_mass}, we show the evolved mass distributions at 100 Myr, for the 30 sets of CIMFs used to reproduce the observed mass distribution reported in CO21 following the same color codes for the simulations as in Fig. \ref{Fig:evol_mass_mayya}, along with their corresponding CIMF.

We show only the two evolved mass distributions at 100 Myr for two CIMFs (labelled as CIMF CO21) to illustrate the mass evolution in terms of the three galactocentric distributions (power-law, exponential disk and uniform) and two surface density distributions, drawn from low and high surface densities. We show in the upper and bottom panels the evolved power-law ($\alpha=2.0)$ and log-normal ($\overline {logM}=4.5$) distributions. In the left panels, we show the less dense whereas in the right panels the denser evolved distributions. Each panel shows the distributions evolving under three different galactocentric initial functions. In the figure, we qualitatively compare the evolved CIMFs with their corresponding CIMF, intended to reproduce the CMF in CO21.

We notice that log-normal CIMFs evolve a log-normal CMFs, being low surface density clusters shifted toward less massive values, than the high surface density ones.  Moreover, clusters evolving under power-law $\rm R_g$ distribution, loose from 0.2 to 0.6 dex. Clusters below $10^4 M_\odot$ are significantly more affected in the low surface density regime, whereas for the high surface initial density, such a limit is $10^{3.4} M_\odot$. On the other hand, power-law distributions show a strong truncation at the low-mass end, departing from the original CIMF trend, with the CIMF evolving under a galactocentric radius power-law distribution, shifted toward lower mass values than the other galactocentric distributions.  The most dramatic change for power-law CIMFs  is given for clusters below $10^{3.4} M_\odot$, with the CIMF evolving from an uniform distribution having larger masses than the other distributions. This trend holds for both low and high surface densities.  For masses above  $10^{3.4} M\odot$ evolving from initial low surface densities, clusters with galactocentric distributions drawn from an uniform distribution, keep the general form of the power-law distribution but shifted one bin toward lower-masses.  This trend is also seen for power-law CIMF with uniform galactocentric  distribution and high surface densities. In the low surface density case, we observe that above $10^{3.4} M\odot$, clusters evolving from galactocentric power-law distributions suffer greater mass-loss, shifting the distributions two bins towards less massive values, differing significantly in the high-mass end (above $10^5 M\odot$) from the other distributions. This is expected, since for initial low-surface densities, clusters are distributed preferentially in zones close to the center, where the galactic field is stronger, favoring strong mass-loss. On the other hand, in the same mass, range, we observe that power-law CIMF, displays similar values for both power-law and exponential disk values, which is due to the smaller radii, causing clusters to be less prone to tidal disruption.

\section{Half-light radius evolution}\label{Sec:ApRh}

In this section, we analyze the evolution of the half-light radius ($\rm R_h$) distribution, in a similar fashion as in Sec. \ref{Sec:mass_ev}, in order to illustrate the $\rm R_h$ evolution in terms of three galactocentric distributions (power-law, exponential disk, and uniform) and two surface density distributions (low and high density), in a similar way as in the previous section for the CMF. As we have mentioned in previous sections, the dataset in M08 does not contain $\rm R_h$ information to be compared with the simulated $\rm R_h$ distributions. 

\begin{figure*}
\begin{center}
\includegraphics[width=0.8\textwidth]{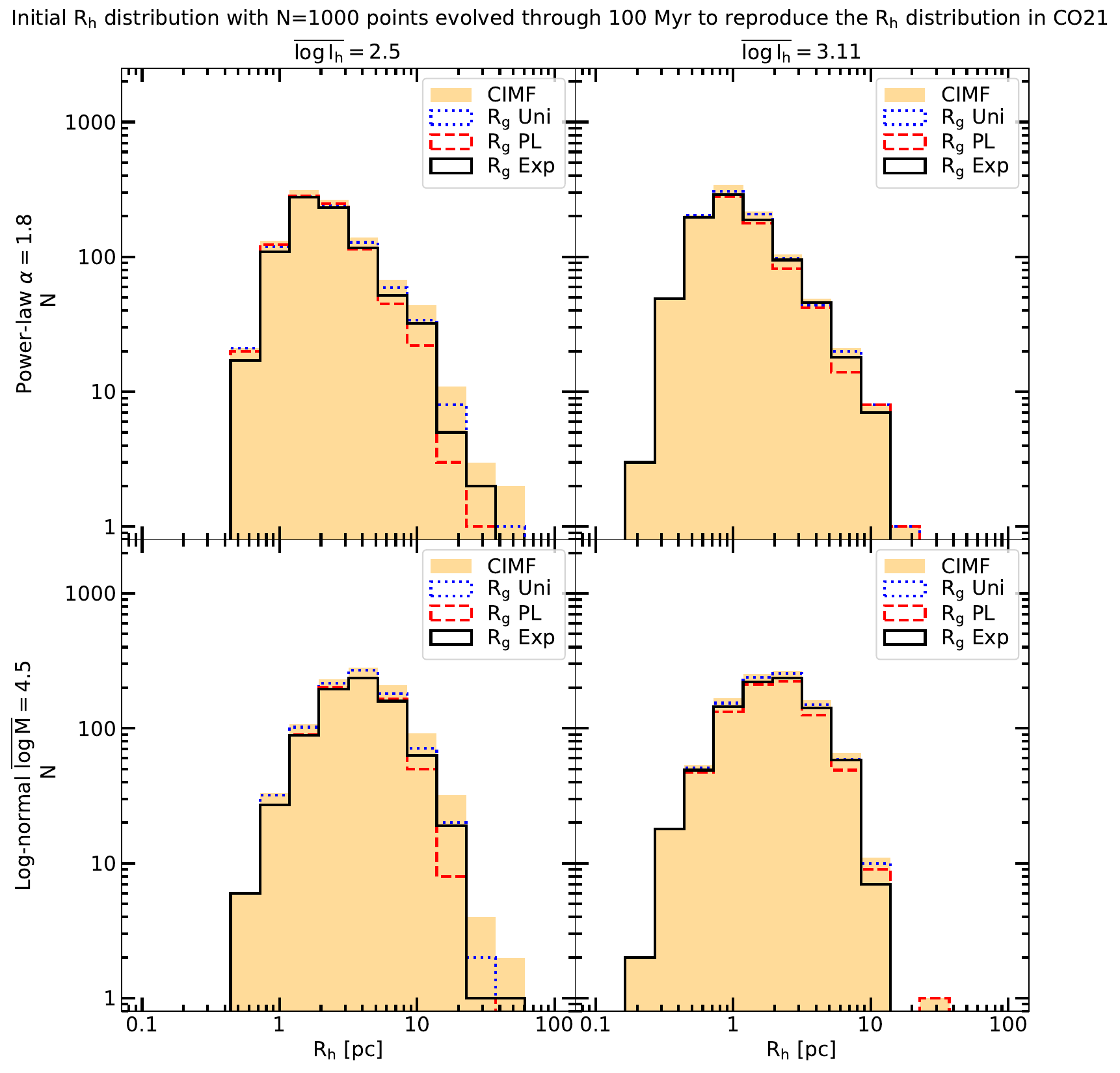}
\caption{$\rm R_h$ distribution evolved through 100 Myr using the semi-analytical evolution code EMACSS. In the left panels, we show $\rm R_h$ distributions evolving assuming clusters with lower surface density distributions, whereas in the right panel, the denser ones. In the top panels we show the evolved $\rm R_h$ distributions with masses drawn from power-law distributions with $\alpha=2$ (runs 7 to 12), whereas in the  bottom panels we show the evolved $\rm R_h$ distributions corresponding to clusters with masses drawn from log-normal distributions with $\overline{\log M=4.5}$ (Runs 13 to 18) under uniform tidal fields (uniform $\rm R_g$ distributions) and exponential and power-law $\rm R_g$.  The evolved $\rm R_h$ distributions are compared with the initial $\rm R_h$ distribution (yellow histogram) drawn from the surface densities distributions along with the CIMs in Figs. \ref{Fig:ini_Ih} and \ref{Fig:ini_mass}, respectively, intended to reproduce the observed $\rm R_h$ distribution in   \citet{Cuevas2020b} (CO21)}
\label{Fig:evol_rad}
\end{center}
\end{figure*}

In Fig. \ref{Fig:evol_rad}, we show the evolution of the half-light radius distribution under different conditions, following the same scheme as Fig.  \ref{Fig:evol_mass}, along with the corresponding initial $\rm R_h$ distributions. As we have pointed out before, the Jacobi radius ($\rm R_j$) dictates the evolution and final fate of a cluster, which is given in terms of the tidal field. In the previous section, we have stressed that the $\rm R_h$ distributions evolving from uniform $\rm R_g$ distributions have on average larger $\rm R_g$, which increases the fraction of larger $\rm R_h$, shifting the distributions toward larger values.   In particular, $\rm R_h$ distributions evolving under log-normal CIMFs conditions have on average larger values.  Log-normal distributions have higher mean mass values (See Tab. \ref{tab:tabla1}) than power-law ones, which explains the larger $\rm R_h$ values since more massive clusters are less prone to be disrupted by tidal effects.  We observe that in general, high surface density distributions, both with power-law and log-normal CIMF, are more similar than their corresponding $\rm R_h$ distributions, with a slight trend of larger clusters ($\rm R_h>7 \, \rm pc$) being disrupted more easily than compact ones. On the other hand, low-surface density clusters, show less extended values, with power-law functions showing the more compact values, due to their smaller $\rm R_j$, due to the combination of shorter $\rm R_g$ and mass loss.   

\section{Observational biases for the sample reported in CO21}

\begin{figure*}
\begin{center}
\includegraphics[width=0.87\textwidth]{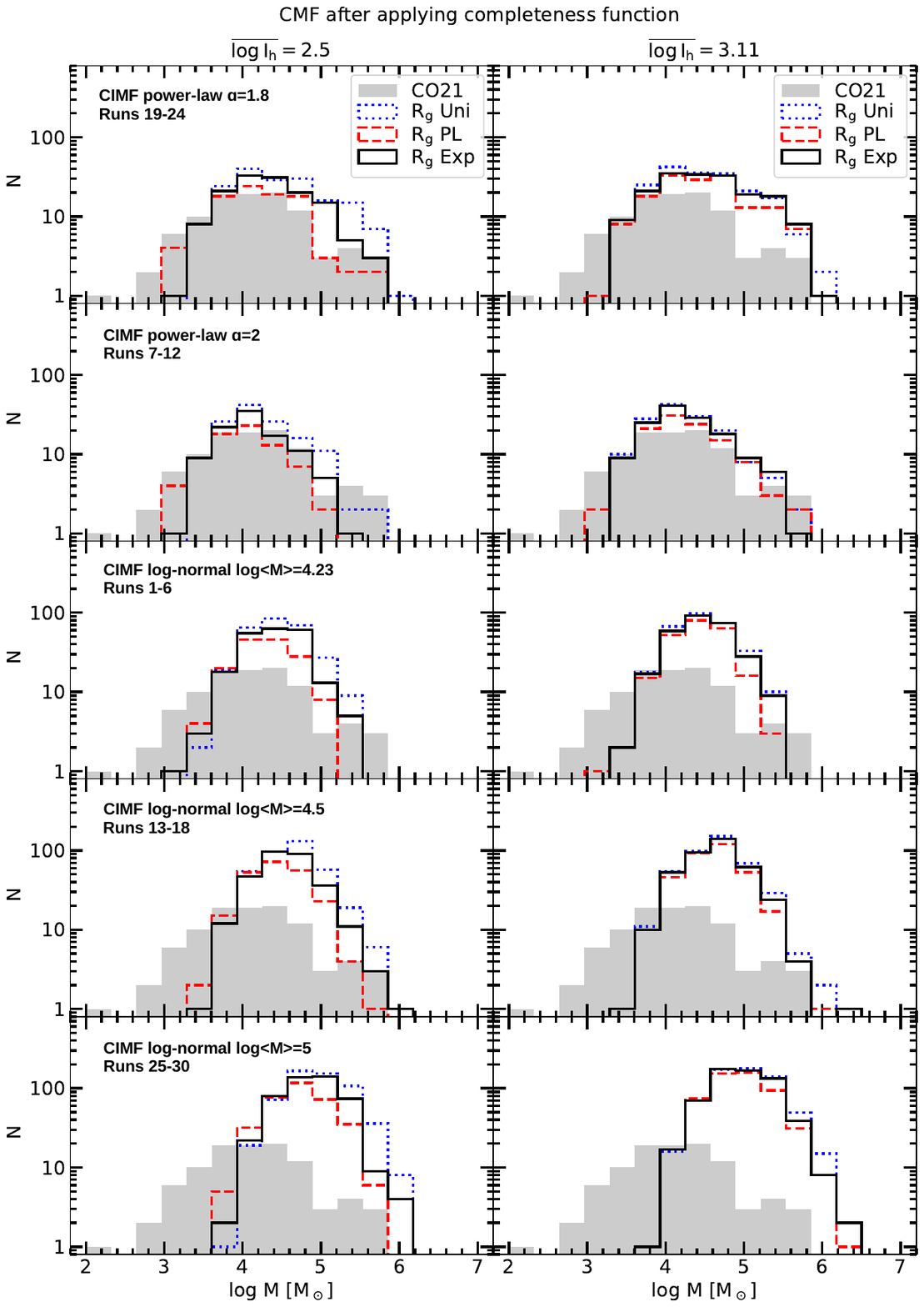}
\caption{CIMFs evolved through 100 Myr using the semi-analytical evolution code EMACSS and applying the completeness function in  Fig. \ref{Fig:comp_func}. In the left panels, we show CIMFs evolving assuming lower surface density distributions, whereas in the right panels, the denser ones. The first and second panels show evolved CIMFs drawn from power-law distributions (Runs 19 to 24, and 7 to 12), and from the middle to the bottom panels evolved CIMFs drawn from log-normal distributions (Runs 1 to 6, 13 to 19, and 25 to 30), under uniform tidal fields (uniform $\rm R_g$ distributions) and exponential and power-law $\rm R_g$.  The evolved CIMFs are compared with the observed CMF in  CO21 (gray histogram).}
\label{Fig:evol_mass_bias}
\end{center}
\end{figure*}

As in Sec. \ref{Sec:obs_eff}, we proceed to apply the completeness function in M08, to the sub-sample CO21, considering that it is a representative subset of the parent distribution in M08, as we have previously laid out. We show these results in Fig. \ref{Fig:evol_mass_bias}. In the middle panel of the figure, we notice that the evolved mass distributions arising from log-normal CIMFs with mean values of 10$^{4.23}\, \rm M_\odot$ do not reproduce the high-mass end of the observed mass distribution.  Also, the fraction of clusters close to the median value is considerably higher than the observed one. These fractions are dramatically larger for log-normal distributions with higher mean values, ruling out these initial conditions, as seen in the middle and bottom panels. In these cases, the resulting biased distributions are shifted toward larger values than the observed distribution. We notice in the first two panels that power-law distributions, on the other hand, reproduce the observed distribution, with the shallower CMF reproducing more accurately the massive end, which is in agreement with the results in Fig. \ref{Fig:evolve_bias_mayya}.

\end{document}